\documentclass[10pt,conference]{IEEEtran}
\IEEEoverridecommandlockouts

\usepackage{cite}
\usepackage{amsmath,amssymb,amsfonts}
\usepackage{graphicx}
\usepackage{textcomp}
\usepackage{xcolor}
\definecolor{Black}{rgb}{0,0,0}
\usepackage{hyperref}
\usepackage{booktabs}
\usepackage{multirow}
\usepackage{enumitem}
\usepackage{algorithm}
\usepackage{algpseudocode}

\newcommand{\new}[1]{{\color{black}#1}}

\newenvironment{keywords}{\begin{center}\bfseries Keywords\end{center}\begin{quote}}{\end{quote}}

\begin{document}

\title{Belief Propagation-Informed Tesseract Decoding of Quantum LDPC Codes
\thanks{The work of S. Javed and M. F. Flanagan was supported by Taighde \'{E}ireann -- Research Ireland under the US-Ireland R\&D Partnership Programme (CoQREATE, Grant Number SFI/21/US-C2C/3750). The work of F. Garcia-Herrero acknowledges support from the project PID2023-147059OB-I00 funded by MCIU/AEI/10.13039/501100011033/FEDER, UE. B. Vasi\'{c} acknowledges support from the NSF under Grants CCF-2100013, CIF-2106189, and CCF-2420424, as well as ERC-1941583 (CoQREATE). A preliminary version of this work was presented at the IEEE International Conference on Quantum Computing and Engineering (QCE), 2026~\cite{javed2026bpinformed}.}}
\author{
\IEEEauthorblockN{Sana Javed$^{\ast}$, 
Francisco Garcia-Herrero$^{\dagger}$, 
Bane Vasi\'{c}$^{\ddagger}$, 
and Mark F. Flanagan$^{\ast}$}

\IEEEauthorblockA{$^{\ast}$School of Electrical and 
Electronic Engineering, University College Dublin, 
Belfield, Dublin 4, Ireland \\
$^{\dagger}$Department of Computer Architecture and 
Automatics, Complutense University of Madrid, 
Madrid, Spain \\
$^{\ddagger}$Department of Electrical and Computer 
Engineering, University of Arizona, 
Tucson, AZ, USA \\
Email: sana.javed@ucdconnect.ie, 
francg18@ucm.es, 
vasic@ece.arizona.edu, 
mark.flanagan@ieee.org}
}

\maketitle

\begin{abstract}
\textcolor{black}{In quantum error correction, the decoder must estimate the errors
from the measurement data accurately and fast enough to avoid a backlog.
Search-based decoders such as Tesseract achieve high accuracy by finding
a minimum-weight set of error events with an A$^*$ search, but their
search effort grows rapidly with the physical error rate and the code
size.} To reduce this search effort while maintaining
\textcolor{black}{a logical error rate close to that} of Tesseract, we propose a
belief propagation (BP)-guided modification of the Tesseract decoder for
quantum \textcolor{black}{low-density parity-check (LDPC) codes, which are
attractive for fault-tolerant quantum computing because of their low
qubit overhead}. Damped sum-product BP is applied to
\textcolor{black}{the triggered detectors} before the Tesseract search. If the
detectors triggered by the BP hard decision \textcolor{black}{on the error vector}
match \textcolor{black}{the triggered detectors}, \textcolor{black}{this hard decision} is accepted directly,
and the Tesseract search is skipped. Otherwise, the BP marginal
probability estimates\textcolor{black}{, weighted by a trust coefficient,} modify
the search priority, while the path cost keeps the channel-prior weights
so that the decoding objective is unchanged. The framework also supports
BP-based initialization, in which selected BP decisions form the
starting point of a residual search whose signed-cost objective allows
incorrect initial decisions to be removed. A node-exploration limit
bounds this residual search; if the limit is reached, the decoder falls
back to the BP-guided search on the original triggered detector set.
The method is evaluated on the
$[\![72,12,6]\!]$, $[\![90,8,10]\!]$, and $[\![144,12,12]\!]$
\new{bivariate bicycle} codes under circuit-level noise in a $Z$-basis
memory experiment with an uncorrelated detector error model. With BP-based initialization and a global trust coefficient, the number of Tesseract search nodes
expanded is reduced by up to more than two orders of magnitude at low
physical error rates, while the logical \textcolor{black}{error rate remains close to
that} of Tesseract.
\end{abstract}


\section{Introduction}
\label{sec:introduction}

Fault-tolerant quantum computers are expected to outperform
classical machines on problems in cryptography,
chemistry, and optimization
\cite{montanaro2016quantum,shor1995scheme,preskill2018quantum,fowler2012surface}.
However, qubits are extremely fragile and prone to errors
caused by decoherence, gate imperfections, and other noise
sources~\cite{nielsenchuang2010,kitaev1997quantum}.
\textcolor{black}{In addition, operations such as non-Clifford gates require the
measurement data to be decoded before the computation can
proceed~\cite{bravyi2005universal,litinski2019magic}.} If measurement data are generated faster than
they can be processed, a backlog can delay the computation~\cite{terhal2015quantum}.

Quantum error correction (QEC) protects against these errors by
encoding each logical qubit into multiple physical qubits;
redundant stabilizer measurements reveal the presence
of errors without disturbing the encoded
information~\cite{terhal2015quantum,nielsenchuang2010}.
Detectors check specified parities of measurement outcomes
and are triggered when these parities differ from their
expected values.
The triggered detector set provides information about the
possible underlying errors.
The detector error model
(DEM)~\cite{gidney2021stim} defines the relationship
between errors and detectors.
One common decoding objective is to find a minimum-weight
error-event set consistent with the triggered detector set.
Under an independent error-event model with channel-prior
log-likelihood weights, this selects the most likely DEM
event configuration consistent with the detector data. It
does not necessarily select the most likely logical class,
whose probability can include contributions from multiple
event configurations.  Finding a minimum-weight error-event set becomes
more computationally demanding as larger circuits and additional
rounds of stabilizer measurements increase the number of error events
and detectors in the DEM.

Quantum LDPC codes reduce qubit overhead while maintaining
good minimum distance, making them attractive for scalable
fault tolerance~\cite{babar2015fifteen}. Efficient decoding
of quantum LDPC codes remains an active research challenge,
particularly due to short cycles in their \textcolor{black}{Tanner graphs}.
Belief propagation (BP) is a computationally attractive
iterative decoder \textcolor{black}{ for quantum LDPC codes}. However, short cycles in the DEM graph
can cause messages to become correlated and oscillate,
so that the detectors triggered by the BP hard decision
\textcolor{black}{on the error vector}
differ from the observed set~\cite{poulin2008iterative}.
As a result, BP-decoded quantum LDPC codes
can exhibit a regime in which the logical
error rate does not decrease as rapidly with the physical
error rate as code distance alone would
predict~\cite{RaveendranVasic2021}.

Several post-processing methods have been proposed to recover
from BP convergence failure. Ordered statistics decoding
(OSD) improves error-correction capability but at the cost of
high latency, limiting parallelization and scalability for
real-time applications~\cite{roffe2020decoding}. Stabilizer inactivation
(SI) offers a lower-complexity alternative by inactivating
unreliable stabilizer nodes and re-running message passing on
the reduced graph~\cite{ducrest2022stabilizer}; however, it
is a list-based method whose complexity grows with the list
length. Serial processing of the list can increase latency,
while parallel processing may demand substantially more
hardware resources~\cite{ducrest2024checkagnosia}.  Localized statistics decoding (LSD)
clusters unreliable qubits and processes them independently
in parallel~\cite{hillmann2025lsd}\textcolor{black}{; however, its cost depends on the
size of the clusters, especially the largest one, and both the cluster
sizes and the decoding time grow as the physical error rate increases}.

An alternative strategy is to modify BP itself.
Belief-propagation-guided decimation (BPGD) sequentially
fixes the most reliable qubits to encourage BP convergence
without post-processing~\cite{yao2024bpgd}, though its
sequential fixing procedure adds latency, and incorrect early
decisions cannot be reversed, unlike search-based methods
that explore multiple error-set candidates. Relay BP~\cite{muller2025relaybp} improves convergence by combining
multiple forms of message damping, achieving competitive decoding
performance on DEM graphs without post-processing\new{, and fully
parallelized BP decoding can also outperform BP-OSD~\cite{wang2026fullyparallelized}}.

Beyond post-processing and BP modifications, the decoding
problem can be embedded in a structured search. The
Tesseract decoder~\cite{beni2025tesseract} and decision-tree
decoders (DTDs)~\cite{Ott2025DecisionTreeDecoders} both
construct \textcolor{black}{error-event sets} incrementally by adding one error at a
time and searching over the space of error subsets. Under the standard assumptions for A$^*$ search, including
nonnegative edge costs, an admissible heuristic, and
appropriate handling of revisited states, an unpruned search
returns a minimum-weight solution.  In practice, Tesseract discards search nodes that leave too
many triggered detectors unexplained and limits the number of
nodes kept in the priority queue. These measures limit the
search effort, so Tesseract operates as a \textcolor{black}{pruned} approximate search.
A recent work~\cite{grbic2026accelerating} accelerated
Tesseract through low-level optimizations such as improved
data structures and cache-friendly memory layouts, reducing
runtime without modifying the search algorithm.
The Tesseract edge weights depend only on the channel priors
and not on the triggered detector set. BP, in contrast,
provides probability estimates conditioned on the triggered
detector set, which can be used to modify the search priority.

\textcolor{black}{BP has also been combined with} search-based
decoders. A BP-based initialization approach proposed by
Yous et al.~\cite{yous2026accelerating} uses a BP hard
decision as an initial error-event set and then performs a
residual search for the detectors that remain unexplained.
\textcolor{black}{In its gated variant, a reliability gate based on the BP
marginal probabilities decides which BP decisions are used; in its
ungated variant, all BP decisions are used.}
\textcolor{black}{In} the BP-DTD variant~\cite{Ott2025DecisionTreeDecoders},
\textcolor{black}{decimated BP is run} at each node of a depth-first search,
using \textcolor{black}{the BP output} to guide local branching decisions.

\textcolor{black}{In this paper,} to reduce the computational cost of search-based decoding,
we propose the \emph{BP-Informed Tesseract} decoder for quantum
LDPC codes under circuit-level noise represented by a DEM.
If the BP hard
decision \textcolor{black}{on the error vector} is consistent with the triggered detector set, it is
accepted directly, avoiding the Tesseract search for that
decoding instance. Otherwise, \textcolor{black}{the BP marginal probabilities}
modify the
search priority, with \new{a trust coefficient
controlling their influence for each error event}. The accumulated path cost remains based on
the original channel-prior weights. BP therefore changes the order
in which the search explores candidates, but not the objective it
minimizes.

\new{In addition to BP guidance, the proposed decoder uses} BP-based initialization for residual
search. Selected error events from the BP hard decision initialize a
residual search that can both add new error events and remove
selected events. \textcolor{black}{When the selected events are correct, fewer
detectors remain unexplained, and the search effort is reduced.}
A signed-cost objective accounts for these
additions and removals\new{: removing an initialized event reduces the
path cost by its channel-prior weight, so the residual search
\textcolor{black}{keeps} the same objective as Tesseract.} A node-exploration limit bounds the effort spent on the
residual search, with fallback to BP-guided Tesseract on the original
triggered detector set if the limit is reached. \new{We consider two
choices of the trust coefficients: a per-error coefficient computed
from the change in each BP marginal probability over the final BP
iterations, and a global coefficient shared by all error events.}

We evaluate the decoder on the $[\![72,12,6]\!]$,
$[\![90,8,10]\!]$, and $[\![144,12,12]\!]$ bivariate bicycle
codes under circuit-level noise.  The experiments use a $Z$-basis memory setting with an
uncorrelated DEM, decode only the $Z$-type detectors,
\textcolor{black}{and examine} the effects of BP guidance,
initialization, and the node-exploration limit. At low physical error rates, \new{the proposed decoder reduces the number of expanded search nodes by \textcolor{black}{up to} more than two orders of magnitude compared with the baseline Tesseract decoder}\textcolor{black}{, while maintaining}
logical error rates close to those of the baseline Tesseract
decoder. \textcolor{black}{A global trust coefficient performs better than the
per-error coefficient on the two larger codes and requires only one
parameter to be selected.}

The remainder of this paper is organized as follows.
Section~\ref{sec:background} introduces quantum LDPC codes, the detector
error model, \textcolor{black}{and the Tesseract and BP decoders}. Section~\ref{sec:method} presents the
proposed decoding method. Section~\ref{sec:results} describes the
experimental setup and presents the numerical results.
Finally, Section~\ref{sec:conclusion} summarizes the main findings and outlines
directions for future work.

\section{Background}
\label{sec:background}

\subsection{Quantum LDPC Codes}
\label{sec:qldpc_codes}

A quantum code with parameters $[\![n,k,d]\!]$ encodes $k$
logical qubits into $n$ physical qubits
and has a minimum distance of $d$.
A quantum LDPC code is a stabilizer code with a set of
stabilizer generators whose weights remain bounded as the
block length increases, with each physical qubit participating
in a bounded number of generators. These generators serve
as the parity checks of the code~\cite{breuckmann2021qldpc,mackay2004good}.
For Calderbank--Shor--Steane (CSS) codes
\cite{calderbank1996good,steane1996multiple},
these generators can be chosen to be either
$X$-type or $Z$-type and are represented by the binary
parity-check matrices $H_X$ and $H_Z$, respectively.
Each row represents the support of a generator, and each
column corresponds to a physical qubit.
The commutativity constraint $H_X H_Z^{T}=0$ over
$\mathbb{F}_2$ ensures that the $X$- and $Z$-type generators
commute.

Quantum LDPC codes can offer a favorable trade-off between
code rate, minimum distance, and qubit overhead, making them
strong candidates for scalable fault-tolerant
architectures~\cite{babar2015fifteen,bravyi2024highthreshold,
breuckmann2021qldpc}.
Several quantum LDPC code families have been
proposed, including hypergraph product
codes~\cite{tillich2013quantum}, lifted product
codes~\cite{panteleev2021degenerate}, and bivariate bicycle (BB)
codes~\cite{bravyi2024highthreshold}.
In this paper, we focus on BB codes, which are CSS codes with $H_X=[A\,|\,B]$
and $H_Z=[B^{T}\,|\,A^{T}]$, where $A$ and $B$ are sums of
products of two commuting cyclic shift
matrices~\cite{bravyi2024highthreshold},
yielding a highly structured sparse code with favorable
distance scaling and low qubit overhead~\cite{kasai2026breaking}.

\subsection{Detector Error Model}
\label{sec:dem}

In quantum error correction, the syndrome is a binary vector of
stabilizer measurement outcomes that provides information about
possible errors~\cite{roffe2020decoding}.
A \emph{detector} checks the parity of syndrome measurement
outcomes, often from consecutive rounds. It is triggered
when this parity differs from the corresponding value in the
absence of faults.
For circuit-level decoding, a DEM
describes how faults in a noisy
syndrome extraction
circuit affect detector outcomes and tracked
logical observables~\cite{gidney2021stim,demarti2026almost}.
The circuit model includes preparation, reset, idling,
ancilla-measurement, and CNOT faults.

Let $E=\{e_1,\ldots,e_N\}$ denote the set of DEM error events.
For notational simplicity, $e_i=1$ denotes the occurrence of
event $e_i$, whereas $e_i=0$ denotes its absence.
The prior probability of event $e_i$ is $p_i=\Pr(e_i=1)$.

Let $\mathcal{D}$ denote the set of detectors and
$S\subseteq\mathcal{D}$ the triggered detector set observed in
one decoding instance. Let $\mathcal{O}=\{o_1,\ldots,o_k\}$
denote the set of $k$ tracked logical observables.
Each event $e_i$ is associated with a detector pattern
$D(e_i)\subseteq\mathcal{D}$ and a logical-observable pattern
$L(e_i)\subseteq\mathcal{O}$.

For a detector $d\in\mathcal{D}$, let
$E(d)=\{e_i\in E:d\in D(e_i)\}$ denote the set of error events
whose detector pattern contains $d$. For an error event set
$F\subseteq E$, the combined detector and logical-observable
patterns are
\begin{equation}
\begin{aligned}
D(F) &= \bigoplus_{e_i\in F}D(e_i)\\
L(F) &= \bigoplus_{e_i\in F}L(e_i)
\end{aligned}
\label{eq:dem_signatures}
\end{equation}
\noindent
where
$\oplus$ denotes symmetric difference. An element belongs
to the resulting pattern if and only if it appears in an
odd number of constituent event patterns. An error event set $F$
is said to be \emph{detector-matching}
when $D(F)=S$. A logical error
occurs when
the decoded logical-observable pattern differs from the
logical-observable pattern generated by the circuit.

\subsection{Tesseract Decoder}
\label{sec:tesseract}

The Tesseract decoder is a search-based decoder for quantum
error correction~\cite{beni2025tesseract,ye2025beam} that seeks a
most-likely error configuration.
Given the triggered detector set $S$, the decoder
searches for a
minimum-weight error configuration consistent with
$S$. Each error event $e_i$ is
assigned a weight based on its associated log-likelihood ratio,
\begin{equation}
    w(e_i)=-\ln\!\left(\frac{p_i}{1-p_i}\right)
    \label{eq:edge_weight}
\end{equation}
and the decoding objective is
\begin{equation}
F^\star \in
\operatorname*{arg\,min}_{\substack{F\subseteq E\\D(F)=S}}
\sum_{e_i\in F}w(e_i)
\label{eq:objective}
\end{equation}
Tesseract formulates this minimum-weight
detector-matching problem
as an A$^*$ search over a directed acyclic graph
(DAG)~\cite{hart1968astar,beni2025tesseract}.
Each node in the graph represents a candidate
error set $F$.
The search begins at the empty error set $F=\emptyset$ and
expands by adding one error event at a time until it reaches a
goal node, which is defined as a node $F$ satisfying $D(F)=S$.
With an admissible heuristic and no pruning, the
first goal node selected for expansion is a solution
of~\eqref{eq:objective}~\cite{hart1968astar}.

Nodes in the search are prioritized by $f(F)=g(F)+h(F)$, where
\begin{equation}
    g(F)=\sum_{e_i\in F}w(e_i)
    \label{eq:path_cost}
\end{equation}
is the accumulated path cost and $h(F)$ is a heuristic
estimate of the remaining cost.
The heuristic is computed from the set of
\textit{residual detectors}
$X(F)=S\oplus D(F)$, whose parity is not matched
by the current partial error set~$F$.

Different orders of adding the same error events can generate
duplicate search paths. Tesseract maintains a forbidden set
$J(F)$
of events excluded from further addition at
node~$F$ to avoid generating the same candidate through different
event orders.

For each detector $d\in X(F)$, the heuristic takes the minimum
ratio of an allowed incident error event's weight to the number of
residual detectors that it triggers:
\begin{equation}
h(F)=
    \sum_{d\in X(F)}
    \min_{e_i\in E(d)\setminus J(F)}
    \frac{w(e_i)}
    {|X(F)\cap D(e_i)|}
    \label{eq:heuristic}
\end{equation}
For nonnegative edge weights,
this heuristic is admissible by construction: the denominator
prevents double-counting, since an error event that flips
multiple unsatisfied detectors has its cost split among
them~\cite{beni2025tesseract}.
\begin{figure*}[h]
    \centering
    \includegraphics[width=\textwidth]{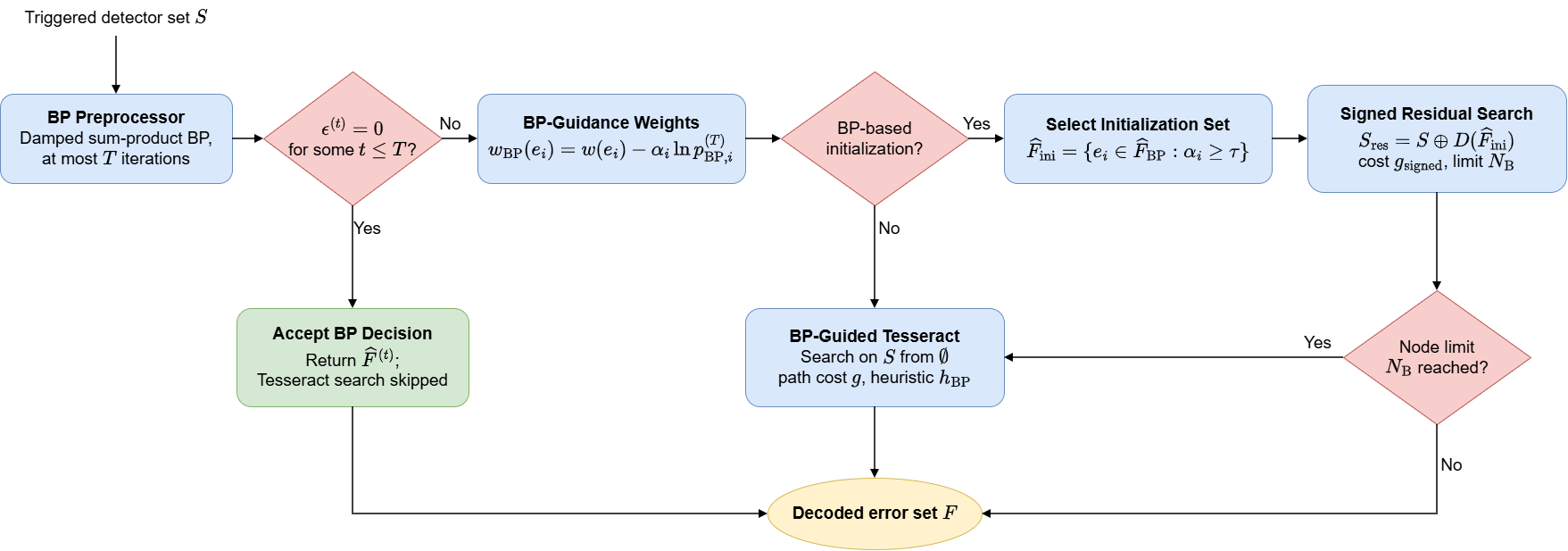}
\caption{Decoder pipeline. If BP finds a detector-matching hard decision
($\epsilon^{(t)}=0$), it is returned directly. Otherwise, the final BP
\textcolor{black}{marginal probabilities} define the BP-informed heuristic.
Without BP-based initialization, BP-guided Tesseract searches on $S$ from the
empty error set. With BP-based initialization, the selected events
$\widehat{F}_{\mathrm{ini}}$ start a residual search; if its node limit
$N_{\mathrm{B}}$ is reached, decoding falls back to BP-guided Tesseract on $S$.
\textcolor{black}{Whether BP-based initialization is used is a configuration
choice made before decoding.}}
\label{fig:pipeline}

\end{figure*}
The search stores candidate nodes in a priority queue and
selects a node with the smallest $f(F)$ for processing.
If it is a goal node,  it is stored as the best solution found so far. Otherwise, the node is
expanded by generating child nodes, each formed by adding
one error event
incident on a selected residual detector
to $F$.
The at-most-two pruning rule rejects \new{any candidate set $F$ in
which some detector is incident on more than two error events of $F$.}
If the queue holds more than $Q_{\max}$ nodes,
priority-queue pruning retains the $Q_{\max}/2$ nodes with the
smallest $f(F)$ values \new{and discards the remaining nodes, which are
never expanded.}

To limit the search further, Tesseract uses a cutoff on the
residual detector count. For a search node $F$, let
$r(F)=|X(F)|$
denote the size of the residual detector set, and
let $r_{\min}$ be the smallest residual detector count encountered
during the search. Given a cutoff parameter $B$, a node is
not expanded if
\begin{equation}
r(F)>r_{\min}+B
\label{eq:tesseract_beam}
\end{equation}
\new{Unlike the forbidden set $J(F)$, which removes only duplicate
paths to the same error set, the} cutoff, the at-most-two rule, and
priority-queue pruning reduce the search effort but can discard
nodes leading to a minimum-weight solution. Thus, the pruned
search is not guaranteed to return a solution
of~\eqref{eq:objective}.


\subsection{Belief Propagation for Quantum Error Correction}
\label{sec:bp}

BP is an iterative message-passing algorithm that computes
exact marginal probabilities
on a tree-structured factor graph
and approximate marginal probabilities when cycles are
present~\cite{poulin2008iterative,kschischang2001factor}.
For the DEM, BP operates on a bipartite graph with two types
of nodes: error-event nodes and detector nodes.
Error-event nodes represent the events $e_i$, detector nodes
represent detector-parity constraints, and an edge connects
$e_i$ to detector $d_j$ when $d_j\in D(e_i)$.
\new{Let $\mathcal{V}$, $\mathcal{C}$, and $\mathcal{E}_{\mathrm{edge}}$
denote the error-event nodes, detector nodes, and edges of this graph.}
The required detector parity at $d_j$ is $s_j$, where $s_j=1$
if $d_j\in S$ and $s_j=0$ otherwise.
Let $\mathcal{N}_{\mathcal{C}}(i)$ denote the set of
detector nodes adjacent to $e_i$, and $\mathcal{N}_{\mathcal{V}}(j)$
the set of error-event nodes adjacent to $d_j$.

Messages are exchanged along the edges in both directions. When sending
a message along an edge to a particular detector, an error-event node
combines its channel prior with the messages from its other neighboring
detectors. A detector node uses
its required parity and messages from the other incident
error-event nodes to update its message to an error-event node.
Messages are expressed as log-likelihood ratios (LLRs).
The channel LLR for each error event $e_i$,
before any detector information is incorporated, is
\begin{equation}
    \lambda_{\mathrm{ch}}(e_i)
    =
    \ln\!\left(\frac{1-p_i}{p_i}\right)
    \label{eq:channel_llr}
\end{equation}
which equals the Tesseract edge weight $w(e_i)$.

\new{BP runs for at most $T$ iterations.} At iteration $t$, let $p_{\mathrm{BP},i}^{(t)}$ denote the BP
estimate of ${\Pr(e_i=1\mid S)}$.
The corresponding marginal LLR is
\begin{equation}
\lambda^{(t)}(e_i)
=
\ln\!\left(
\frac{1-p_{\mathrm{BP},i}^{(t)}}
     {p_{\mathrm{BP},i}^{(t)}}
\right)
\label{eq:posterior_llr}
\end{equation}

The hard decision at iteration $t$ is
\begin{equation}
\widehat{F}^{(t)}
=
\left\{
e_i\in E:\lambda^{(t)}(e_i)<0
\right\}
\label{eq:bp_hard_decision}
\end{equation}

The number of residual detectors for this hard decision is
\begin{equation}
    \epsilon^{(t)}
    =
    \left|D\!\left(\widehat{F}^{(t)}\right)\oplus S\right|
    \label{eq:residual_mismatch}
\end{equation}

Thus, $\epsilon^{(t)}=0$ means that the BP hard decision
$\widehat{F}^{(t)}$ is detector-matching, i.e.,
$D(\widehat{F}^{(t)})=S$. We say that BP \emph{converges} if
$\epsilon^{(t)}=0$ for some $t\le T$; BP then stops and returns
$\widehat{F}^{(t)}$. This criterion does not require the messages to reach
a fixed point, and a detector-matching hard decision need not be
logically correct.

Although BP is effective for many classical LDPC codes,
short cycles in quantum LDPC graphs
can cause messages to become correlated
and make the resulting marginal probability
estimates inaccurate. BP may therefore fail to converge, oscillate
between belief states, or produce a hard decision that does not
match the triggered detector
set~\cite{poulin2008iterative,RaveendranVasic2021}.

Damping combines a newly computed BP message with its value
from the previous iteration and has been shown to improve
convergence in loopy BP~\cite{Sibel2013}.
Giving more weight to the previous messages can reduce
oscillations, although it may slow convergence.

\section{Proposed Method: BP-Informed Tesseract}
\label{sec:method}

\textcolor{black}{
The proposed decoder integrates a damped BP preprocessor with the
Tesseract search. Its key idea is to use the BP output to reduce the
Tesseract search effort while keeping the channel-prior path cost
unchanged. For a given triggered detector set $S$, BP is run first. If
its hard decision is detector-matching at any iteration, it is returned
as the decoded error set and the Tesseract search is skipped.
Otherwise, the final BP marginal probabilities modify the Tesseract
heuristic, so that error events that BP considers likely are expanded
earlier. A trust coefficient $\alpha_i$ controls the influence of BP on
each error event $e_i$; a global coefficient corresponds to the same
value for all error events.
}

\textcolor{black}{
The search can also start from the BP output instead of the empty error
set. With BP-based initialization, error events selected from the final
BP hard decision form an initialization set, and Tesseract searches only
for the detectors that this set leaves unexplained; we refer to this as
the residual search. Since some selected events may be incorrect, the
residual search uses a signed-cost objective that allows initialized
events to be removed while keeping the channel-prior objective. A
node-exploration limit $N_{\mathrm{B}}$ bounds the residual search; if
it is reached, the decoder falls back to BP-guided Tesseract on $S$.
The overall decoding procedure is illustrated in
Fig.~\ref{fig:pipeline}.
}
\subsection{BP Preprocessor}
\label{sec:bp_preprocessor}
\textcolor{black}{For the given} triggered detector set $S$, \new{the decoder runs damped BP}
on the bipartite DEM graph. For each edge $(i,j)\in\mathcal{E}_{\mathrm{edge}}$,
$\mu_{i\rightarrow j}^{(t)}$ denotes the message from error event $e_i$
to detector node $d_j$, and $\nu_{j\rightarrow i}^{(t)}$ denotes the
message in the reverse direction at iteration $t$.
\new{Algorithm~\ref{alg:bp_preprocessor} \textcolor{black}{describes} the damped BP
preprocessor.} \textcolor{black}{Its message updates and damping are those of
standard damped sum-product BP (Section~\ref{sec:bp}); it is
included here to define the stopping rule and the outputs used by the
subsequent stages of the decoder.} \new{In Algorithm~\ref{alg:bp_preprocessor},}
$\widetilde{\mu}_{i\rightarrow j}^{(t)}$ is the
undamped error-to-detector message, $\beta$ is the damping factor,
$\boldsymbol{\lambda}^{(t)}=\{\lambda^{(t)}(e_i)\}_{i=1}^{N}$ denotes the marginal LLRs
at iteration $t$, and
$\mathcal{P}^{(t)}=\{p_{\mathrm{BP},i}^{(q)}: i=1,\ldots,N,\ q=1,\ldots,t\}$
denotes the marginal probability history up to iteration $t$.

\begin{algorithm}[h]
\caption{\new{Damped} BP Preprocessor}
\label{alg:bp_preprocessor}
\begin{algorithmic}[1]

    \State \textbf{Input:} Triggered detector set $S$,
           channel LLRs $\{\lambda_{\mathrm{ch}}(e_i)\}_{i=1}^{N}$,
           DEM bipartite graph
           $\mathcal{G}=(\mathcal{V}\cup\mathcal{C},
           \mathcal{E}_{\mathrm{edge}})$,
           damping factor $\beta\in(0,1]$, and maximum number of
           iterations $T$

    \State \textbf{Output:} BP hard decision
           $\widehat{F}_{\mathrm{BP}}$, \new{marginal LLRs
           $\boldsymbol{\lambda}^{(t)}$, marginal probability history
           $\mathcal{P}^{(t)}$}, and success flag
           $\mathrm{flag}_{\mathrm{BP}}\in\{0,1\}$

    \State \textbf{Initialize:}
           $\mu_{i\rightarrow j}^{(0)}
           \leftarrow\lambda_{\mathrm{ch}}(e_i)$
           and $\nu_{j\rightarrow i}^{(0)}\leftarrow0$
           for every edge $(i,j)\in\mathcal{E}_{\mathrm{edge}}$

    \State $\mathrm{flag}_{\mathrm{BP}}\leftarrow0$

    \For{$t=1$ \textbf{to} $T$}

        \ForAll{$(i,j)\in\mathcal{E}_{\mathrm{edge}}$}
\State \resizebox{0.87\linewidth}{!}{$\nu_{j\rightarrow i}^{(t)} \leftarrow (-1)^{s_j}\,
       2\tanh^{-1}\!\bigl(\prod_{i'\in\mathcal{N}_{\mathcal{V}}(j)\setminus\{i\}}
       \tanh(\mu_{i'\rightarrow j}^{(t-1)}/2)\bigr)$}
        \EndFor

        \ForAll{$(i,j)\in\mathcal{E}_{\mathrm{edge}}$}
            \State $\widetilde{\mu}_{i\rightarrow j}^{(t)}
                   \leftarrow\lambda_{\mathrm{ch}}(e_i)
                   +
                   \sum_{j'\in
                   \mathcal{N}_{\mathcal{C}}(i)\setminus\{j\}}
                   \nu_{j'\rightarrow i}^{(t)}$

            \State $\mu_{i\rightarrow j}^{(t)}
                   \leftarrow
                   \beta\widetilde{\mu}_{i\rightarrow j}^{(t)}
                   +(1-\beta)\mu_{i\rightarrow j}^{(t-1)}$
        \EndFor

        \ForAll{$i\in\{1,\ldots,N\}$}
            \State $\lambda^{(t)}(e_i)
                   \leftarrow\lambda_{\mathrm{ch}}(e_i)
                   +
                   \sum_{j\in\mathcal{N}_{\mathcal{C}}(i)}
                   \nu_{j\rightarrow i}^{(t)}$

            \State $p_{\mathrm{BP},i}^{(t)}
                   \leftarrow
                   [1+\exp(\lambda^{(t)}(e_i))]^{-1}$
        \EndFor

        \State $\widehat{F}^{(t)}
               \leftarrow
               \{e_i\in E:\lambda^{(t)}(e_i)<0\}$

        \State $\epsilon^{(t)}
               \leftarrow
               \left|
               D(\widehat{F}^{(t)})\oplus S
               \right|$

        \If{$\epsilon^{(t)}=0$}

            \State $\widehat{F}_{\mathrm{BP}}
                   \leftarrow\widehat{F}^{(t)}$

            \State $\mathrm{flag}_{\mathrm{BP}}\leftarrow1$

            \State \Return $\widehat{F}_{\mathrm{BP}}$, $\boldsymbol{\lambda}^{(t)}$, $\mathcal{P}^{(t)}$, $\mathrm{flag}_{\mathrm{BP}}$

        \EndIf

    \EndFor

    \State $\widehat{F}_{\mathrm{BP}}
           \leftarrow\widehat{F}^{(T)}$

    \State \Return $\widehat{F}_{\mathrm{BP}}$, $\boldsymbol{\lambda}^{(T)}$, $\mathcal{P}^{(T)}$, $\mathrm{flag}_{\mathrm{BP}}$

\end{algorithmic}
\end{algorithm}

If $\mathrm{flag}_{\mathrm{BP}}=1$, the detector-matching decision
$\widehat{F}_{\mathrm{BP}}$ is \new{returned as the decoded error set},
although it need not be a minimum-weight detector-matching set.

\subsection{BP Weight Guidance}
\label{sec:weight_guidance}

When BP does not produce a detector-matching hard decision within $T$
iterations, the final marginal probabilities
$p_{\mathrm{BP},i}^{(T)}$ are used to modify the Tesseract heuristic.
\textcolor{black}{For each error event $e_i$, a trust coefficient
$\alpha_i\in[0,1]$ controls the influence of its estimate.}

We define the BP-guidance weight as
\begin{equation}
    w_{\mathrm{BP}}(e_i)
    =
    w(e_i)
    -
    \textcolor{black}{\alpha_i}\ln p_{\mathrm{BP},i}^{(T)}
    \label{eq:bp_weight}
\end{equation}
where $w(e_i)$ is the channel-prior weight defined
in~\eqref{eq:edge_weight}.

For $0<p_{\mathrm{BP},i}^{(T)}<1$, the correction term
$-\textcolor{black}{\alpha_i}\ln p_{\mathrm{BP},i}^{(T)}$
is nonnegative:
for \textcolor{black}{$\alpha_i>0$},
it increases $w_{\mathrm{BP}}(e_i)$ for \new{error events} that BP considers
unlikely.
This correction guides the search:
\new{error events} that BP identifies as unlikely are penalized in the expansion
order \new{through a larger correction term}, while \new{error events} that BP
identifies as likely receive a smaller \new{correction term}.

We substitute $w_{\mathrm{BP}}(e_i)$ for $w(e_i)$ in the heuristic
of~\eqref{eq:heuristic}, giving the BP-informed
heuristic $h_{\mathrm{BP}}(F)$.
The path cost $g(F)$ continues to use
the channel-prior weights.
The search priority is therefore
\begin{equation}
    f_{\mathrm{BP}}(F)
    =
    g(F)+h_{\mathrm{BP}}(F)
    \label{eq:bp_search_priority}
\end{equation}
\new{We refer to the Tesseract search with search priority
$f_{\mathrm{BP}}$ as BP-guided Tesseract.}
Thus, BP changes the search order while
the decoding objective remains the original channel-prior cost.

Under the nonnegative-weight condition for which $h(F)$ is admissible,
and using the same residual detector set and allowed events,
$w_{\mathrm{BP}}(e_i)\geq w(e_i)$ implies
$h_{\mathrm{BP}}(F)\geq h(F)$. Thus, $h_{\mathrm{BP}}$ may overestimate
the remaining channel-prior cost. Admissibility is therefore not
guaranteed for \textcolor{black}{$\alpha_i>0$}.

\subsection{Trust Coefficients}
\label{sec:alpha}

\new{The BP marginal probabilities} may continue to change across
iterations even with damping\textcolor{black}{, and they can behave
differently for different error events: some may settle, while others
may continue to change or oscillate. To capture this variation, each
error event $e_i$ has its own trust coefficient $\alpha_i$
in~\eqref{eq:bp_weight}.} Each coefficient is computed directly from the
\new{marginal probability history $\mathcal{P}^{(T)}$}, without separate
tuning for individual error events.

\textcolor{black}{Following the mean-change statistic of Yous
\emph{et al.}~\cite{yous2026accelerating}, we measure the mean absolute
change in the marginal probability of $e_i$ over the last $W$ BP
iterations:}
\begin{equation}
\Delta_{\mathrm{mean},i}
=\frac{1}{W}\sum_{t=T-W}^{T-1}
\left|p_{\mathrm{BP},i}^{(t+1)}-p_{\mathrm{BP},i}^{(t)}\right|
\label{eq:delta_mean}
\end{equation}
\textcolor{black}{The per-error coefficient is then defined as}
\begin{equation}
\textcolor{black}{\alpha_i}
=
\frac{1}{1+\Delta_{\mathrm{mean},i}}
\label{eq:alpha_i}
\end{equation}
Since $\Delta_{\mathrm{mean},i}\in[0,1]$,
$\textcolor{black}{\alpha_i}\in[0.5,1]$. A marginal  probability that changes very
little has $\textcolor{black}{\alpha_i}\approx1$, whereas one that continues to change receives a smaller coefficient.
\textcolor{black}{We use this form because it decreases with
$\Delta_{\mathrm{mean},i}$, equals the largest value when the marginal
probability does not change, keeps at least half of the BP correction
for every error event, and requires no additional parameter.}
This coefficient \textcolor{black}{measures how much the marginal
probability changes}. A stable but incorrect
\new{BP marginal probability} can therefore receive a large coefficient.

\textcolor{black}{A special case is the global coefficient, in which the
same value $\alpha_i=\alpha$ is used for all error events.}
A larger $\alpha$
gives the BP marginal probabilities greater influence
on the search priority and can reduce the number of node expansions
when the \new{BP marginal probabilities} are accurate. A
smaller $\alpha$ keeps the heuristic closer to the
original channel-prior heuristic, reducing its dependence on
potentially unreliable \new{BP marginal probabilities}.
The value of $\alpha$ is selected \new{empirically}
\textcolor{black}{by comparing several values, and we use $\alpha=0.5$ for all codes.
This value is the lower bound of the per-error coefficient
in~\eqref{eq:alpha_i}, corresponding to $\Delta_{\mathrm{mean},i}=1$.
Unlike the per-error coefficient, the global coefficient applies this
value to every error event, regardless of the observed change in its
marginal probability, so marginal probabilities that are stable but
incorrect do not receive a larger coefficient. It also requires only
one parameter to be selected.}
\subsection{BP-Based Initialization for Residual Search}
\label{sec:init}

The trust-coefficient mechanism described above guides the search;
it does not alter the channel-prior-based path cost or restrict the
candidate error sets.

BP-based initialization uses the BP output more directly, by starting
the search from selected BP decisions rather than from the empty error
set. We use a subset of the final BP hard decision
$\widehat{F}_{\mathrm{BP}}$ to
initialize a residual search. \textcolor{black}{The} initialization set is
\begin{equation}
\widehat{F}_{\mathrm{ini}}
    =
    \left\{
    e_i\in\widehat{F}_{\mathrm{BP}}:
    \textcolor{black}{\alpha_i}\geq\tau
    \right\}
    \label{eq:tau_gate}
\end{equation}
where $\tau\in[0,1]$ is a selection threshold.
An error event is therefore included in the
initialization only if it is present in the BP hard decision and its
\textcolor{black}{coefficient $\alpha_i$} reaches the threshold.
With $\tau=0$, all events of $\widehat{F}_{\mathrm{BP}}$
are selected, as in the ungated initialization of Yous
\emph{et al.}~\cite{yous2026accelerating}\new{; the proposed decoder
differs \textcolor{black}{from that of~\cite{yous2026accelerating}} in the signed-cost objective and the node-exploration
limit}.

The residual search then starts with the residual detector set
\begin{equation}
S_{\mathrm{res}}
    =
    S\oplus D\!\left(\widehat{F}_{\mathrm{ini}}\right)
    \label{eq:residual_detector_set_warm}
\end{equation}

The set $S_{\mathrm{res}}$ contains the detectors whose parity is not
matched by the selected initialization. To illustrate how a correct
initialization can simplify the remaining decoding problem, let
$F_{\mathrm{true}}\subseteq E$ denote the underlying error-event set in
the DEM representation, so that $D(F_{\mathrm{true}})=S$. This set is
introduced only for analysis and is not known to the decoder. If
$\widehat{F}_{\mathrm{ini}}\subseteq F_{\mathrm{true}}$, then, because
$D$ is linear over $\mathbb{F}_2$,
\begin{equation}
S_{\mathrm{res}}
    =
    S\oplus D\!\left(\widehat{F}_{\mathrm{ini}}\right)
    =
    D\!\left(F_{\mathrm{true}}\setminus\widehat{F}_{\mathrm{ini}}\right)
\end{equation}
Under this condition, the residual detector set is
produced by the remaining \new{error events}. The residual search can therefore
require less search effort than a search starting from the original
triggered detector set.

If a selected event is not in $F_{\mathrm{true}}$, the residual search
can remove it.
A lower threshold can include more error events
in $\widehat{F}_{\mathrm{ini}}$, whereas a higher threshold can
include fewer. If no error event satisfies the selection condition,
the initialization set is empty and $S_{\mathrm{res}}=S$.
\subsection{Signed-Cost Objective for the Residual Search}
\label{sec:signedwarm}

For the residual detector set $S_{\mathrm{res}}$ defined
in~\eqref{eq:residual_detector_set_warm}, the residual search seeks a
residual correction
$\Delta F$ satisfying
\begin{equation}
D(\Delta F)=S_{\mathrm{res}}
\label{eq:residual_constraint}
\end{equation}
The final error set is
\begin{equation}
F=\widehat{F}_{\mathrm{ini}}\oplus\Delta F
\label{eq:final_estimate}
\end{equation}
\new{\textcolor{black}{The residual correction $\Delta F$ may contain events of $\widehat{F}_{\mathrm{ini}}$.}
By~\eqref{eq:final_estimate},} an event
in $\Delta F\setminus\widehat{F}_{\mathrm{ini}}$ is added
to the initial estimate, while an event in
$\Delta F\cap\widehat{F}_{\mathrm{ini}}$ is removed \new{from it}. By the
linearity of $D$ over $\mathbb{F}_2$, a residual correction
satisfying~\eqref{eq:residual_constraint} gives
\begin{align}
D(F)
&=
D\!\left(\widehat{F}_{\mathrm{ini}}\right)\oplus D(\Delta F)
\nonumber\\
&=
D\!\left(\widehat{F}_{\mathrm{ini}}\right)\oplus S_{\mathrm{res}}
=S
\label{eq:final_detector_consistency}
\end{align}

More than one residual correction can satisfy~\eqref{eq:residual_constraint}. We
compare them using the \new{channel-prior cost $w(F)$ of the final error
set, which is the objective of~\eqref{eq:objective} and equals $g(F)$
in~\eqref{eq:path_cost}}. Since the DEM models
error events as independent, $w(F)$ and $-\ln P(F)$ differ only by a
constant independent of $F$, where
\new{$P(F)=\prod_{e_i\in F}p_i\prod_{e_i\notin F}(1-p_i)$ is the
probability of the error-event set $F$ under the DEM.} Minimizing
$w(F)$ over detector-matching error sets therefore selects a
most-likely error-event set.

Substituting $F=\widehat{F}_{\mathrm{ini}}\oplus\Delta F$ gives
\begin{align}
w(F)
&=
w\!\left(\widehat{F}_{\mathrm{ini}}\right)
-\sum_{e_i\in\Delta F\cap\widehat{F}_{\mathrm{ini}}}w(e_i)
+\sum_{e_i\in\Delta F\setminus\widehat{F}_{\mathrm{ini}}}w(e_i)
\nonumber\\
&=
w\!\left(\widehat{F}_{\mathrm{ini}}\right)
+\sum_{e_i\in\Delta F}\sigma_i w(e_i)
\label{eq:signed_cost}
\end{align}
where
\begin{equation}
\sigma_i=
\begin{cases}
+1, & e_i\notin\widehat{F}_{\mathrm{ini}},\\
-1, & e_i\in\widehat{F}_{\mathrm{ini}}
\end{cases}
\label{eq:sigma_def}
\end{equation}
\textcolor{black}{By~\eqref{eq:signed_cost}, removing an initialized event $e_i$
changes $w(F)$ by $-w(e_i)$. A residual search with nonnegative edge
costs would instead charge $+w(e_i)$ for this removal, overestimating
its cost by $2w(e_i)$ and favoring solutions that retain incorrect
initialized events.}
We therefore use the signed residual path cost
\begin{equation}
g_{\mathrm{signed}}(\Delta F)
=
w(F)-w\!\left(\widehat{F}_{\mathrm{ini}}\right)
=
\sum_{e_i\in\Delta F}\sigma_i w(e_i)
\label{eq:signed_objective}
\end{equation}

Since $w(\widehat{F}_{\mathrm{ini}})$ \textcolor{black}{does not depend on $\Delta F$},
minimizing $g_{\mathrm{signed}}$ over $\Delta F$ is equivalent to
minimizing $w(F)$.
The mapping $F=\widehat{F}_{\mathrm{ini}}\oplus\Delta F$ is a
bijection between residual corrections satisfying
$D(\Delta F)=S_{\mathrm{res}}$ and final error sets satisfying
$D(F)=S$.
Over all residual corrections
satisfying~\eqref{eq:residual_constraint}, let
\begin{equation}
\Delta F^\star \in
\operatorname*{arg\,min}_{D(\Delta F)=S_{\mathrm{res}}}
g_{\mathrm{signed}}(\Delta F)
\label{eq:signed_ml_equivalence}
\end{equation}
Then
$F^\star=\widehat{F}_{\mathrm{ini}}\oplus\Delta F^\star$
solves~\eqref{eq:objective}, and every solution of
\eqref{eq:objective} corresponds to a minimizer of
\new{$g_{\mathrm{signed}}$} under this mapping.
\textcolor{black}{The residual search runs the Tesseract search on $S_{\mathrm{res}}$
with path cost $g_{\mathrm{signed}}$ and the BP-informed heuristic to seek a solution of~\eqref{eq:signed_ml_equivalence}.}
\begin{algorithm}[t]
\caption{Unified BP-Informed Decoder}
\label{alg:unified}
\begin{algorithmic}[1]
\State \textbf{Input:} $S$, $\{w(e_i)\}$, $\mathcal{G}$, $\beta$, $T$, $W$,
       $\alpha$, $\tau$, $N_{\mathrm{B}}$; guidance (global or
       \textcolor{black}{per-error}); with or without BP-based initialization
\State \textbf{Output:} Decoded error set $F$ or decoding failure
\State $(\widehat{F}_{\mathrm{BP}},\boldsymbol{\lambda}^{(t)},\mathcal{P}^{(t)},
       \mathrm{flag}_{\mathrm{BP}})\leftarrow$
\Statex \hspace{\algorithmicindent}\textcolor{black}{Algorithm~\ref{alg:bp_preprocessor}$(S,\{\lambda_{\mathrm{ch}}(e_i)\},\mathcal{G},\beta,T)$}
\If{$\mathrm{flag}_{\mathrm{BP}}=1$}
    \State \Return $\widehat{F}_{\mathrm{BP}}$
\EndIf
\State $\alpha_i\leftarrow\alpha$ (global) or
       \textcolor{black}{$\alpha_i$ from~\eqref{eq:alpha_i}}
\State $w_{\mathrm{BP}}(e_i)\leftarrow w(e_i)-\alpha_i\ln p_{\mathrm{BP},i}^{(T)}$
       \Comment{Weights for $h_{\mathrm{BP}}$}
\If{BP-based initialization is not used}
    \State \Return BP-guided Tesseract on $S$ from $\emptyset$
\EndIf
\State $\widehat{F}_{\mathrm{ini}}\leftarrow\{e_i\in\widehat{F}_{\mathrm{BP}}:
       \textcolor{black}{\alpha_i}\geq\tau\}$
\State $S_{\mathrm{res}}\leftarrow S\oplus D(\widehat{F}_{\mathrm{ini}})$
\State Run the residual search on $S_{\mathrm{res}}$ with
       $g_{\mathrm{signed}}$, $h_{\mathrm{BP}}$, and limit $N_{\mathrm{B}}$
\If{the limit $N_{\mathrm{B}}$ is reached}
    \State \Return BP-guided Tesseract on $S$ from $\emptyset$
\ElsIf{a residual correction $\Delta F$ is found}
    \State \Return $\widehat{F}_{\mathrm{ini}}\oplus\Delta F$
\Else
    \State \Return failure (no detector-matching set found)
\EndIf
\end{algorithmic}
\end{algorithm}
\textcolor{black}{Since $\sigma_i=-1$ for initialized events, removing such an event
has a negative edge cost. Each initialized event can be removed at most
once along a search path, so the path cost can decrease by at most
$w(\widehat{F}_{\mathrm{ini}})$; if no initialized event is removed, all
edge costs are nonnegative.} 

\new{The \textcolor{black}{pruned} search is not
guaranteed to find a global minimum. \textcolor{black}{In addition,} because
$g_{\mathrm{signed}}$ permits negative increments, the BP-informed
heuristic is not guaranteed to be a lower bound on the remaining signed
cost.}

\subsection{Node-Limited Search}
\label{sec:nodecap}

\new{If some selected BP decisions are incorrect, the residual search
must remove them through $\Delta F$, which can require many node
expansions.} We limit each
residual-search attempt to $N_{\mathrm{B}}$ expanded nodes.
If this limit is reached before the residual search
terminates, its result is discarded and the decoder falls back to
BP-guided Tesseract on the original triggered detector set $S$,
starting from the empty error set.

The cutoff parameter $B$, defined in~\eqref{eq:tesseract_beam}, serves
a different purpose: it restricts which nodes can be expanded according
to the size of their residual detector set. In contrast, $N_{\mathrm{B}}$
limits the number of expansions in the residual-search attempt. \new{Algorithm~\ref{alg:unified} summarizes the complete decoding
procedure.}

\begin{table}[t]
\centering
\caption{DEM parameters \textcolor{black}{of the three BB codes}.}
\label{tab:circuit_dims}
\begin{tabular}{lcccc}
\toprule
Code & Rounds & Detectors & Error events & Observables \\
\midrule
$[\![72,12,6]\!]$   & 6  & 252 & 2{,}232 & 12 \\
$[\![90,8,10]\!]$   & 10 & 495 & 4{,}590 & 8  \\
$[\![144,12,12]\!]$ & 12 & 936 & 8{,}784 & 12 \\
\bottomrule
\end{tabular}
\end{table}

\section{Results}
\label{sec:results}

We evaluate the proposed decoder on the $[\![72,12,6]\!]$,
$[\![90,8,10]\!]$, and $[\![144,12,12]\!]$ \new{BB}
codes from~\cite{bravyi2024highthreshold}. For each code, circuits are
generated using \textsc{Stim}~\cite{gidney2021stim} for a $Z$-basis
memory experiment\textcolor{black}{, in which the logical qubits are prepared and
measured in the $Z$ basis,} with \new{$r=d$} syndrome-extraction rounds\new{.} The
resulting \new{DEM} dimensions are
summarized in Table~\ref{tab:circuit_dims}.

We use the circuit-level depolarizing noise model in \textsc{Stim} and
simulate physical error rates from $p=0.002$ to \new{$0.007$} in steps of
$0.001$. \new{For a fair comparison, all decoders are evaluated on the
same detector samples.} The normalized logical error rate per round is
\begin{equation}
    \text{LER}_{\text{round}}
    = \frac{N_{\text{logical}}}{N_{\text{shots}}\times \new{r}}
    \label{eq:ler_round}
\end{equation}
where $N_{\mathrm{shots}}$ is the number of \new{sampled} shots and
$N_{\mathrm{logical}}$ is the number of shots in which the decoded
observable flips differ from the sampled ones. Error bars \new{at all
points} show 95\% Wilson \textcolor{black}{score} confidence intervals\textcolor{black}{,
computed for $N_{\mathrm{logical}}/N_{\mathrm{shots}}$ and divided by $r$}.
\new{In addition to the LER per round, we report} the average number of
Tesseract nodes expanded per shot, and \textcolor{black}{the maximum priority-queue
size reached during the search of a shot, averaged over all shots}. A node is counted each time it is removed
from the priority queue. \textcolor{black}{For decoders with a node-exploration
limit, the nodes expanded in an abandoned residual-search attempt are
included in the node count.}

The Tesseract search uses a cutoff on the
residual detector count with $B=20$\new{, the value recommended
in~\cite{beni2025tesseract},} priority-queue pruning with
$Q_{\max}=5000$, and the at-most-two rule\new{~\cite{beni2025tesseract}}.
\new{The value of $Q_{\max}$ is chosen to keep the simulation time of
the larger codes manageable.} These pruning settings are shared by all
decoders. Queued nodes whose search-priority values exceed
\textcolor{black}{the cost of the best detector-matching error set found so far}
are discarded\new{, and} the search ends when the queue is empty.

The BP preprocessor uses the sum-product algorithm with a flooding
schedule. The values of $\beta$, $\alpha$, $N_{\mathrm{B}}$, $T$, and
$W$, along with the configuration-specific values of $\tau$, are
described in the next subsection.
\begin{figure}[t]
\centering
\includegraphics[width=0.99\columnwidth]{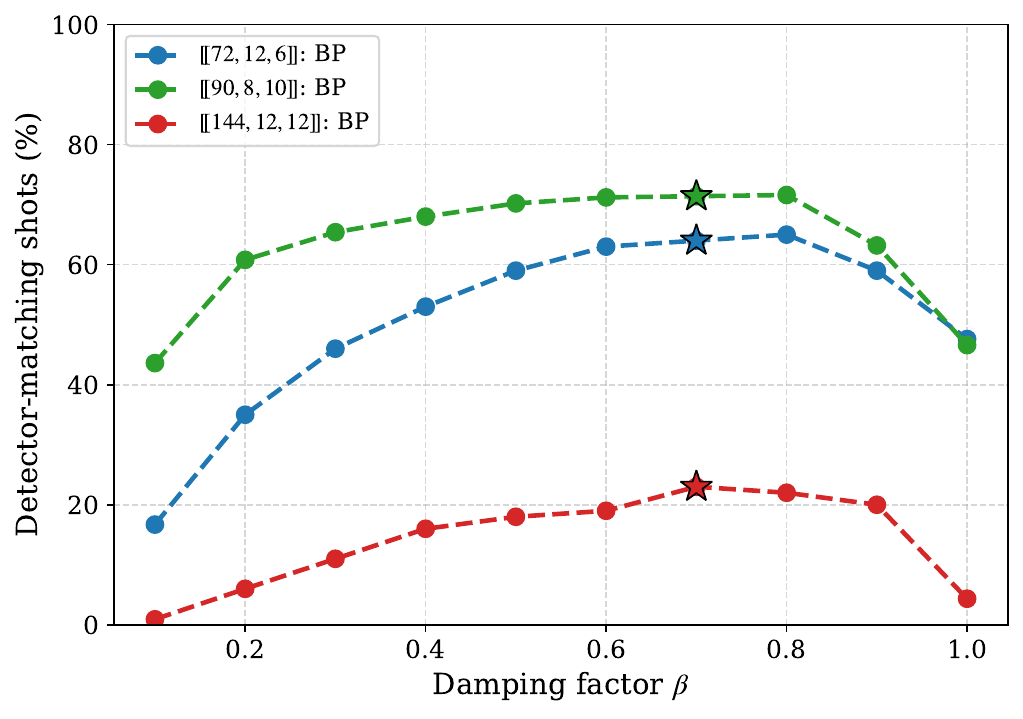}
\caption{BP convergence rate versus damping factor
         $\beta$ at $p = 0.005$ for the
         $[\![72,12,6]\!]$, $[\![90,8,10]\!]$, and $[\![144,12,12]\!]$ BB
         codes. Stars mark the adopted default $\beta = 0.7$ for each
         code.}
\label{fig:beta_ablation}
\end{figure}
\textcolor{black}{We compare the Tesseract baseline, the proposed decoder and its
variants, and the ungated initialization method
of~\cite{yous2026accelerating}}, named as in the figures. \emph{Tesseract
(A$^*$)} is the baseline search without BP. \emph{BP-informed guidance}
\new{is the BP-guided Tesseract decoder:} a detector-matching BP hard
decision is accepted directly; otherwise, the BP \new{marginal
probabilities} with $\alpha=0.5$ guide the Tesseract search. The \emph{BP-based init.} decoders start a residual search from the
selected BP decisions and use the signed-cost objective. Each attempt
is limited to $N_{\mathrm{B}}=50$ node expansions. If the limit is reached, the shot is re-decoded with BP-informed
guidance. \emph{Global} uses the \textcolor{black}{global}
coefficient $\alpha$ and $\tau=0$, whereas \emph{\textcolor{black}{per-error}}\ uses the
\textcolor{black}{per-error coefficients $\alpha_i$ from~\eqref{eq:alpha_i}} with
$\tau=0.9$. \emph{Ungated init.} is the ungated variant of the BP-based initialization
of Yous~\emph{et al.}~\cite{yous2026accelerating}, which initializes from
all events of the BP hard decision, uses edge weights computed from the
BP marginal probabilities for both the path cost and the heuristic, and
has neither the signed-cost objective nor a node limit.
\subsection{Parameter Selection}
\label{sec:parameter_selection}
We first select the damping factor.
Fig.~\ref{fig:beta_ablation} shows the fraction of shots for which BP
produces a detector-matching hard decision as a function of
\new{$\beta$}. The results are obtained at
$p=0.005$ using $10{,}000$ shots for each value of $\beta$ and each
code. \new{We select $\beta=0.7$, which increases the fraction of detector-matching
shots compared with undamped BP ($\beta=1$) for all three codes, and use
the same value for all codes and physical error rates. We fix the maximum number of BP
iterations to $T=50$ and the \textcolor{black}{window $W$} to the last 20\% of the
BP iterations, i.e., $W=10$. For the $[\![72,12,6]\!]$ code at
$p=0.005$, shots for which BP produces a detector-matching hard decision
require on average $7.8$ iterations (median $5$), well below $T=50$.}
\begin{table}[t]
\centering
\caption{Effect of the global trust coefficient $\alpha$ for the
$[\![72,12,6]\!]$ code at $p=0.005$.}
\label{tab:alpha_selection}
\renewcommand{\arraystretch}{1.2}
\begin{tabular}{@{}c c c@{}}
\toprule
$\alpha$ & Average nodes expanded & LER per round \\
\midrule
0.0 & 617.0 & $1.33\times10^{-2}$ \\

0.2 & 154.1 & $1.33\times10^{-2}$ \\

0.4 & 69.2  & $1.17\times10^{-2}$ \\
\textbf{0.5} & \textbf{51.2} & $\mathbf{1.08\times10^{-2}}$ \\
0.6 & 42.0  & $1.17\times10^{-2}$ \\

0.8 & 33.4  & $1.25\times10^{-2}$ \\

1.0 & 27.3  & $1.33\times10^{-2}$ \\
\bottomrule
\end{tabular}
\end{table}

Next, we select the global trust coefficient $\alpha$, which controls
how strongly the BP \new{marginal probabilities} modify the search
heuristic when BP does not produce a detector-matching hard decision.
We evaluate \new{the values of $\alpha$ listed in
Table~\ref{tab:alpha_selection}}. The case $\alpha=0$ provides a
zero-correction reference corresponding to the original Tesseract
heuristic. As shown in Table~\ref{tab:alpha_selection}, increasing
$\alpha$ generally reduces the number of expanded nodes, but the
measured logical error rate is not monotonic.
\textcolor{black}{Compared with $\alpha=0$, which expands $617.0$ nodes on average,
$\alpha=0.5$ expands $51.2$ nodes} while \new{keeping} a measured logical error rate
comparable to the other tested values. We therefore select $\alpha=0.5$
using the $[\![72,12,6]\!]$ code at $p=0.005$ and hold it fixed for the
remaining codes and physical error rates.

\begin{table}[h]
\centering
\caption{\new{BP-based init.\ (global)} with node-exploration limits
$N_{\mathrm{B}}=50$ and $N_{\mathrm{B}}=200$: logical error rate per
round (LER) and average number of nodes expanded per shot\textcolor{black}{, averaged
over all shots, including the shots with a detector-matching BP hard
decision, for which no node is expanded}.}
\label{tab:nodelimit}
\renewcommand{\arraystretch}{1.15}
\setlength{\tabcolsep}{3pt}
\begin{tabular}{@{}llcccc@{}}
\toprule
 & & \multicolumn{2}{c}{$N_{\mathrm{B}}=50$} & \multicolumn{2}{c}{$N_{\mathrm{B}}=200$} \\
\cmidrule(lr){3-4}\cmidrule(lr){5-6}
Code & $p$ & LER & Nodes & LER & Nodes \\
\midrule
\multirow{2}{*}{$[\![72,12,6]\!]$}
 & 0.003 & $1.38\times10^{-3}$ & 1.29 & $1.44\times10^{-3}$ & 1.04 \\
 & 0.007 & $3.97\times10^{-2}$ & 475 & $4.08\times10^{-2}$ & 436 \\
\midrule
\multirow{2}{*}{$[\![90,8,10]\!]$}
 & 0.003 & $1.30\times10^{-4}$ & 8.85 & $1.77\times10^{-4}$ & 7.64 \\
 & 0.007 & $2.29\times10^{-2}$ & 3{,}158 & $2.19\times10^{-2}$ & 3{,}036 \\
\midrule
\multirow{2}{*}{$[\![144,12,12]\!]$}
 & 0.003 & $3.17\times10^{-5}$ & 10.4 & $3.67\times10^{-5}$ & 8.52 \\
 & 0.007 & $2.26\times10^{-2}$ & 8{,}855 & $2.19\times10^{-2}$ & 8{,}689 \\
\bottomrule
\end{tabular}
\end{table}
For the configuration with
global $\alpha$, we set $\tau=0$, so all error events in the final BP
hard decision satisfy the selection condition and form the
initialization set. For the \textcolor{black}{per-error} configuration, we
use $\tau=0.90$ for all three codes \textcolor{black}{ without further tuning}.
\textcolor{black}{Since $\alpha_i\geq0.90$ in~\eqref{eq:alpha_i} requires
$\Delta_{\mathrm{mean},i}\leq1/0.90-1\approx0.11$,} an event is selected only if its marginal
probability changes by at most $0.11$ per iteration on average
over the \textcolor{black}{last $W$ BP iterations}.

Finally, we select the node-exploration limit. The limit bounds the
residual search started from the BP initialization. A smaller value
therefore activates the fallback \new{to BP-guided Tesseract} earlier,
whereas a larger value allows more node expansions in the residual
search before falling back. Table~\ref{tab:nodelimit} compares
\new{BP-based init.\ (global)} with $N_{\mathrm{B}}=50$ and
$N_{\mathrm{B}}=200$, using the same global coefficient $\alpha$ and
selection threshold $\tau=0$. The larger limit \new{slightly reduces} the average node count\new{,
while at $p=0.003$ the measured logical error rate is slightly higher,
most noticeably for the $[\![90,8,10]\!]$ code.} We therefore use
$N_{\mathrm{B}}=50$ as the default.
\subsection{Global and Per-Error Trust Coefficients}
\label{sec:per_error_results}
Table~\ref{tab:per_error} compares \new{BP-informed guidance} using the
global coefficient $\alpha$ with the \textcolor{black}{per-error
coefficient $\alpha_i$}.

For the $[\![72,12,6]\!]$ code, the two configurations have comparable
measured logical error rates, while the per-error configuration expands
fewer nodes than the global configuration at both tested values of $p$.
For the highest tested physical error rate, $p=0.007$, the global and
per-error configurations expand $519$ and $196$ nodes per shot,
respectively.

The per-error configuration does not show the same improvement for the
two larger codes. For the $[\![90,8,10]\!]$ code, the per-error
configuration has a higher measured logical error rate than the global
configuration at both tested values of $p$\new{, while its node count is
similar at $p=0.003$ and lower at $p=0.007$}. For $p=0.007$, the global
and per-error configurations expand \new{$3{,}282$ and $2{,}325$} nodes
per shot, respectively. For the
$[\![144,12,12]\!]$ code at the same physical error rate, the
corresponding values are \new{$9{,}761$} and $22{,}526$ nodes per shot,
and the per-error configuration also has a higher measured logical error
rate.
\begin{table}[t]
\centering
\caption{Logical error rate per round (LER) and average number of nodes
expanded per shot for Tesseract and for BP-informed guidance with the
global coefficient $\alpha$ and with the \textcolor{black}{per-error
coefficient $\alpha_i$}.}
\label{tab:per_error}
\renewcommand{\arraystretch}{1.15}
\setlength{\tabcolsep}{3pt}
\begin{tabular}{@{}llcccc@{}}
\toprule
 & & \multicolumn{2}{c}{$p=0.003$} & \multicolumn{2}{c}{$p=0.007$} \\
\cmidrule(lr){3-4}\cmidrule(lr){5-6}
Code & Decoder & LER & Nodes & LER & Nodes \\
\midrule
\multirow{3}{*}{$[\![72,12,6]\!]$}
 & Tesseract    & $1.30\times10^{-3}$ & 51.5 & $4.10\times10^{-2}$ & \new{3{,}883} \\
 & \new{Global} & $1.20\times10^{-3}$ & 3.03 & $3.88\times10^{-2}$ & 519 \\
 & \new{Per-error} & $1.25\times10^{-3}$ & 2.81 & $4.03\times10^{-2}$ & 196 \\
\midrule
\multirow{3}{*}{$[\![90,8,10]\!]$}
 & Tesseract    & $1.13\times10^{-4}$ & 273.7 & $2.31\times10^{-2}$ & 15{,}803 \\
 & \new{Global} & $1.17\times10^{-4}$ & 11.0  & $2.39\times10^{-2}$ & \new{3{,}282} \\
 & \new{Per-error} & $1.60\times10^{-4}$ & 11.1  & $2.67\times10^{-2}$ & \new{2{,}325} \\
\midrule
\multirow{3}{*}{$[\![144,12,12]\!]$}
 & Tesseract    & $3.83\times10^{-5}$ & \new{1{,}043} & $2.17\times10^{-2}$ & 38{,}702 \\
 & \new{Global} & $3.17\times10^{-5}$ & 17.1 & $2.29\times10^{-2}$ & \new{9{,}761} \\
 & \new{Per-error} & $7.33\times10^{-5}$ & 22.1 & $2.74\times10^{-2}$ & 22{,}526 \\
\bottomrule
\end{tabular}
\end{table}
\begin{figure*}[t]
\centering
\includegraphics[width=0.98\textwidth,height=6cm]
{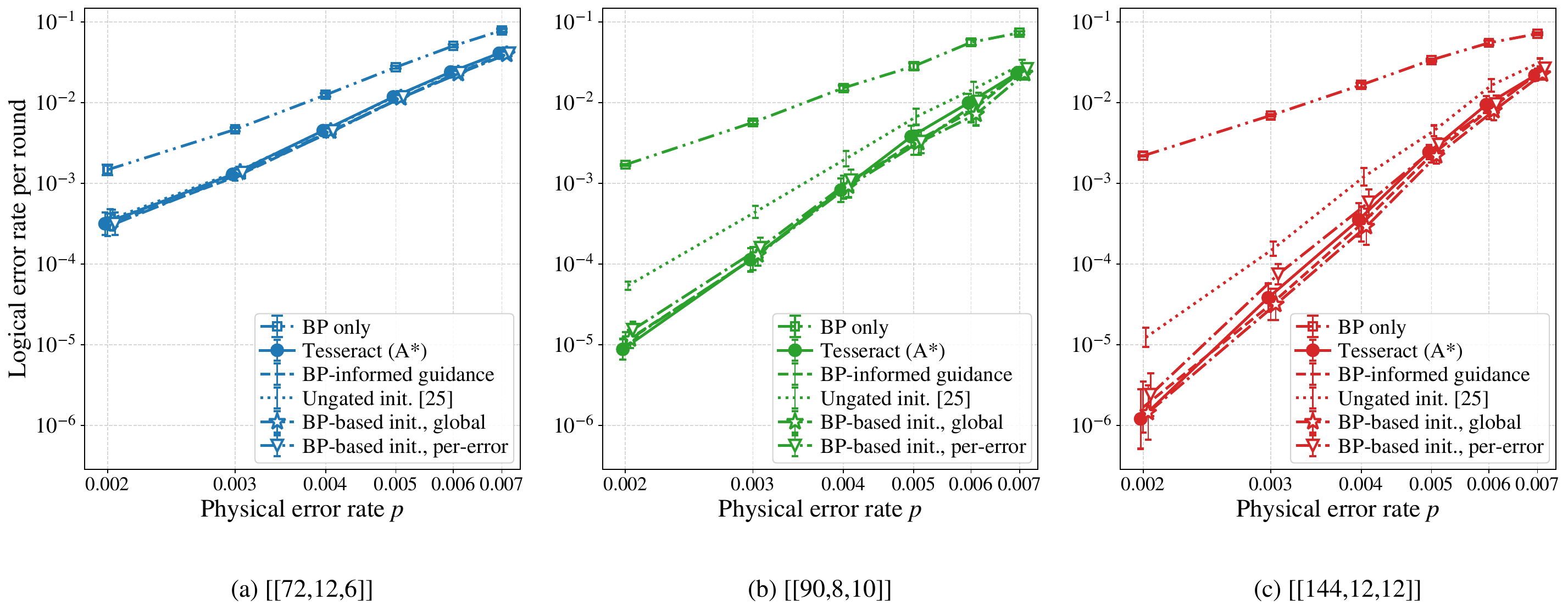}

\caption{Logical error rate per round versus physical error rate $p$ for
the $[\![72,12,6]\!]$, $[\![90,8,10]\!]$, and $[\![144,12,12]\!]$ codes.
Error bars show 95\% Wilson \textcolor{black}{score} confidence intervals; for damped BP alone, they are smaller than the markers. The comparison includes
damped BP alone ($\beta = 0.7$), the Tesseract baseline, BP-informed guidance, the ungated
\textcolor{black}{variant of the} initialization
of~\cite{yous2026accelerating}, and the proposed
BP-based initialization with the \new{signed-cost objective} and
$N_{\mathrm{B}}=50$. All configurations use $B=20$.}

\label{fig:ler_full}
\end{figure*}
The per-error coefficient gives each error event its own weight correction. A large \textcolor{black}{per-error} coefficient indicates that a
\new{marginal probability} is stable, but does not guarantee that the
preferred BP decision is correct. For the \new{$[\![90,8,10]\!]$ and
$[\![144,12,12]\!]$ codes, whose DEMs contain $4{,}590$ and $8{,}784$
error events,} the per-error coefficient may place incorrectly favored
error events earlier in the search order. This provides one possible
explanation for the higher measured logical error rates.
A global coefficient works better \textcolor{black}{on the two larger codes} and requires only one parameter to be selected. We
therefore use the global coefficient as the main configuration and report the per-error coefficient as a supplementary
comparison.

\subsection{Logical Error Rate \new{per Round}}
\label{sec:ler}

Fig.~\ref{fig:ler_full} shows the logical error rate per round as a
function of the physical error rate $p$ for the three codes. \textcolor{black}{The
comparison includes damped BP alone, the Tesseract baseline, the proposed decoder and its
variants, and the ungated initialization method
of~\cite{yous2026accelerating}, which we evaluate under circuit-level noise.}

For the $[\![72,12,6]\!]$ code, BP-informed guidance\textcolor{black}{, the proposed
decoder, and its per-error variant} give measured logical error rates close to the
Tesseract baseline throughout the tested range. For this code, the BP output therefore does not cause a noticeable loss
in accuracy.

For the $[\![90,8,10]\!]$ code, the decoders differ more clearly in
their logical error rates. The ungated
initialization has a higher measured logical error rate than the
Tesseract baseline throughout the tested range: at $p=0.002$, its
measured value is approximately $6.2$ times the baseline value
($5.42\times10^{-5}$ versus $8.80\times10^{-6}$), and the ratio decreases
to approximately $1.29$ at $p=0.007$. \new{\textcolor{black}{One possible reason is
that, without} the signed-cost
objective, \textcolor{black}{removing an incorrect initialized event is charged
$+w(e_i)$ instead of $-w(e_i)$}, which biases the search toward
keeping incorrect initialization events.} In
contrast, the \textcolor{black}{proposed decoder remains} close to the
baseline over the tested range.

For the $[\![144,12,12]\!]$ code, the same pattern is observed, with
lower logical error rates overall due to the larger code distance. The
Tesseract baseline decreases from $2.17\times10^{-2}$ at $p=0.007$ to
$3.83\times10^{-5}$ at $p=0.003$, a much steeper decrease than for the
$[\![72,12,6]\!]$ code, consistent with the stronger suppression
expected at higher distance. The ungated initialization again has a
higher measured logical error rate, approximately $4.0$ times the
baseline value at $p=0.003$ and $1.46$ times at $p=0.007$.
BP-based initialization with global $\alpha$ remains close to the
baseline, whereas BP-based initialization with \textcolor{black}{the per-error
coefficient} has a higher measured logical error rate \new{ at low $p$}. \new{For this
code,} at $p=0.002$, only a few logical
failures are observed, so the corresponding confidence intervals are
wide.

For comparison, Fig.~\ref{fig:ler_full} also shows damped BP alone, where
a shot is counted as a failure if BP does not converge or converges to a
correction with a wrong logical observable. BP alone has a much higher
logical error rate than all search-based decoders. At $p=0.002$, it is
about $5$ times that of Tesseract for the $[\![72,12,6]\!]$ code and two
to three orders of magnitude higher for the two larger codes, mainly
because BP does not converge on these shots.

\begin{table}[t]
\centering
\caption{Percentage of shots on which BP converges (accepted directly,
no Tesseract search) for each code and physical error rate $p$.}
\label{tab:bp_branches}
\renewcommand{\arraystretch}{1.1}
\begin{tabular}{@{}lccc@{}}
\toprule
$p$ & $[\![72,12,6]\!]$ & $[\![90,8,10]\!]$ & $[\![144,12,12]\!]$ \\
\midrule
0.002 & 99.3 & 98.3 & 97.4 \\
0.003 & 97.6 & 94.3 & 91.7 \\
0.004 & 93.5 & 84.8 & 80.1 \\
0.005 & 85.2 & 71.7 & 59.1 \\
0.006 & 72.1 & 44.0 & 33.9 \\
0.007 & 56.0 & 26.5 & 14.3 \\
\bottomrule
\end{tabular}
\end{table}
\begin{figure*}[t]
\centering
\includegraphics[width=0.98\textwidth]
{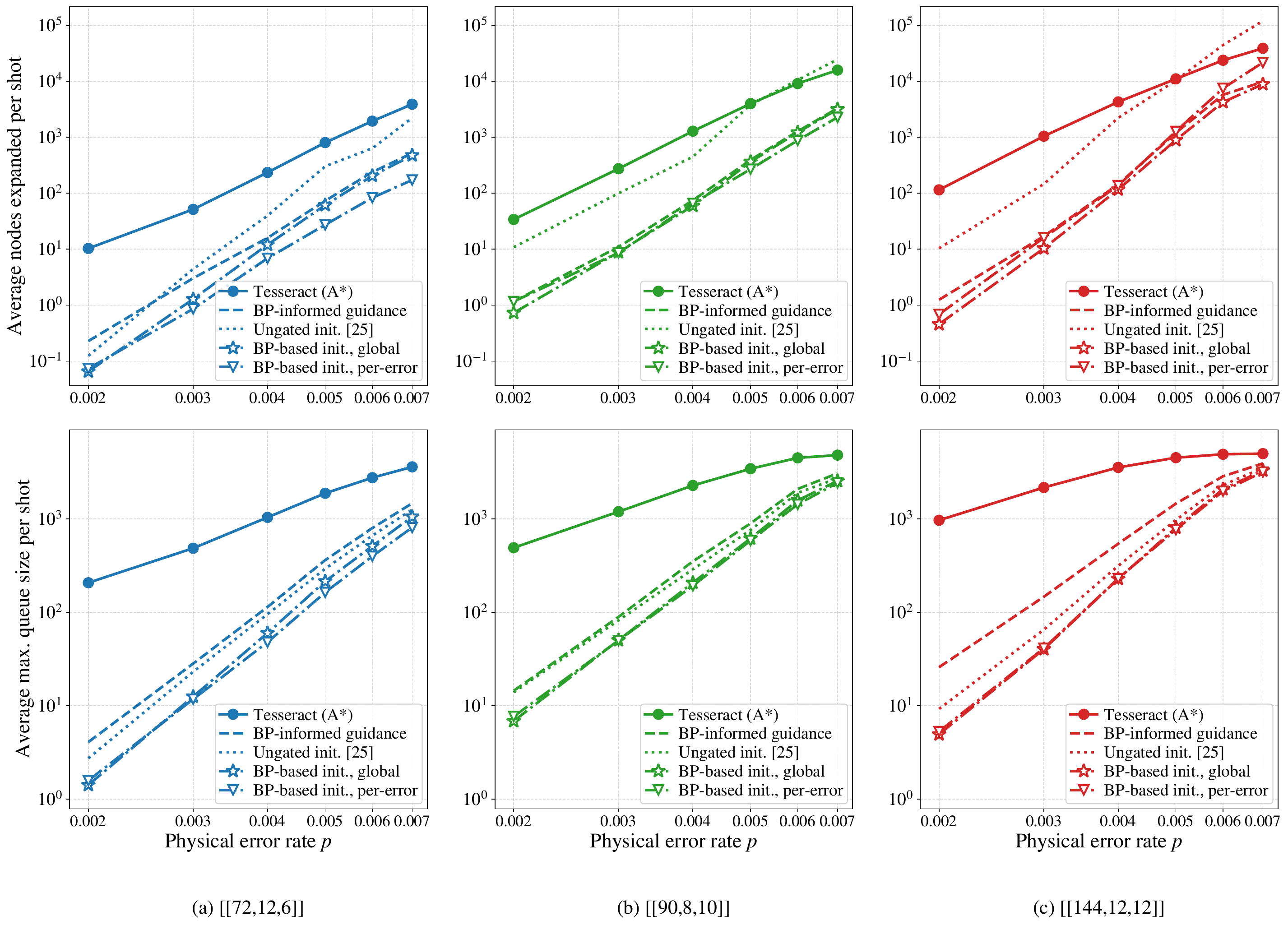}
\caption{Average number of nodes expanded per shot (top row) and
\textcolor{black}{maximum priority-queue size reached during the search of a
shot, averaged over all shots} (bottom row) versus physical error
rate $p$. Columns correspond to the $[\![72,12,6]\!]$,
$[\![90,8,10]\!]$, and $[\![144,12,12]\!]$ codes. Here, ``global''
denotes one coefficient $\alpha$ for all error events, while
``\textcolor{black}{per-error}'' denotes the \textcolor{black}{per-error coefficient
$\alpha_i$}. All BP-based initialization
configurations use the \new{signed-cost objective and $N_B=50$}.}
\label{fig:nodes_queue_6panel}
\end{figure*}
\subsection{Average Number of Nodes Expanded}
\label{sec:avg_nodes}

The top row of Fig.~\ref{fig:nodes_queue_6panel} shows the average
number of nodes expanded per shot over all shots, including nodes
expanded during both the residual-search attempt and any subsequent
fallback \textcolor{black}{ to BP-guided Tesseract}. \new{Table~\ref{tab:bp_branches} gives the percentage of
shots on which BP converges, so that the Tesseract search is skipped.
The convergence rate decreases as $p$ and the code size increase: at
$p=0.002$, BP converges on 99.3\%, 98.3\%, and 97.4\% of the shots of
the $[\![72,12,6]\!]$, $[\![90,8,10]\!]$, and $[\![144,12,12]\!]$
codes, respectively, whereas at $p=0.007$ it converges on only 56.0\%,
26.5\%, and 14.3\%.}

For the $[\![72,12,6]\!]$ code, BP-based initialization with global
$\alpha$ and $N_{\mathrm{B}}=50$ reduces the average from \new{$3{,}883$}
nodes for the Tesseract baseline to $475$ at $p=0.007$, an $8.2\times$
reduction. At $p=0.002$, BP-informed guidance reduces the average from
$10.3$ to $0.23$ nodes, and BP-based initialization further reduces it to
$0.066$ nodes, an overall reduction of approximately $157\times$. This is
consistent with BP converging, and the Tesseract search being skipped, on
the \textcolor{black}{vast} majority of shots at $p=0.002$ (Table~\ref{tab:bp_branches}).

For the $[\![90,8,10]\!]$ code, BP-based initialization with global
$\alpha$ reduces the average from $15{,}803$ to \new{$3{,}158$} nodes at
$p=0.007$, a $5.0\times$ reduction, and from $34.0$ to $0.73$ nodes at
$p=0.002$, a $46\times$ reduction.

For the $[\![144,12,12]\!]$ code, the node counts are substantially
larger in absolute terms, reflecting the larger DEM in
Table~\ref{tab:circuit_dims}: at $p=0.007$, the Tesseract baseline
expands $38{,}702$ nodes per shot, approximately $10\times$ more than for
the $[\![72,12,6]\!]$ code. BP-based initialization with global
$\alpha$ reduces this to \new{$8{,}855$} nodes, a $4.4\times$ reduction,
and reduces the average from \new{$114.5$} to \new{$0.45$} nodes at
$p=0.002$, an approximately \new{$254\times$} reduction.

\new{At $p=0.002$, BP-based initialization with global $\alpha$ reduces
the node count relative to BP-informed guidance by $3.5\times$,
$1.6\times$, and $2.8\times$ for the $[\![72,12,6]\!]$,
$[\![90,8,10]\!]$, and $[\![144,12,12]\!]$ codes, respectively.}
For all three codes, the reduction factor increases as the physical
error rate decreases.

The ungated initialization behaves differently. It expands fewer nodes
than the Tesseract baseline at low $p$, but more for $p\ge0.006$ on the
$[\![90,8,10]\!]$ and $[\![144,12,12]\!]$ codes; at $p=0.007$, it
expands $1.6\times$ and $3.1\times$ more nodes, respectively. Because this configuration has no additional node-exploration limit,
it can spend more search effort on difficult shots, even though the shared cutoff and priority-queue pruning remain active.

\subsection{Maximum Priority-Queue Size}
\label{sec:avg_queue}

The bottom row of Fig.~\ref{fig:nodes_queue_6panel} shows
\textcolor{black}{the maximum priority-queue size reached during the search of a
shot, averaged over all shots}. This quantity is a proxy for the
peak memory demand of the search, which is relevant for hardware
implementations with limited buffer size. For shots on which the
node-exploration limit $N_{\mathrm{B}}$ is reached, we take the larger
of the maximum queue sizes reached during the residual and fallback
searches.

For the $[\![72,12,6]\!]$ code, BP-based initialization with global
$\alpha$ maintains a smaller queue than the Tesseract baseline at every
tested rate: \textcolor{black}{this quantity} drops from \new{$3{,}620$} to
\new{$1{,}059$} at $p=0.007$, and from $208$ to $1.42$ at $p=0.002$.
For the two larger codes, the queue sizes are considerably larger,
consistent with their larger DEMs. At $p=0.007$, the reduction is
smaller, $1.9\times$ and $1.5\times$ for the $[\![90,8,10]\!]$ and
$[\![144,12,12]\!]$ codes, possibly because
there are many competing low-weight candidates at this rate. Here the Tesseract baseline queue is close to
$Q_{\max}$, so its reported size underestimates that of \new{a search
without priority-queue pruning}. As $p$ decreases, the reduction grows,
reaching $72\times$ and $197\times$ at $p=0.002$.
\new{At $p=0.002$, BP-based initialization reduces \textcolor{black}{this quantity}
relative to BP-informed guidance by $2.9\times$, $2.1\times$,
and $5.3\times$ for the $[\![72,12,6]\!]$, $[\![90,8,10]\!]$, and
$[\![144,12,12]\!]$ codes, respectively.} \textcolor{black}{The per-error coefficient} gives
queue sizes similar to those of the global $\alpha$.

\textcolor{black}{In summary, with} global $\alpha$, BP-based initialization keeps the measured logical error rate close to Tesseract and reduces both the number of expanded nodes and the queue size. The reductions are largest at low $p$, where BP converges on most shots.
\section{Conclusion}
\label{sec:conclusion}

\textcolor{black}{We have proposed BP-Informed Tesseract, a decoder for quantum LDPC
codes that uses the BP output to reduce the Tesseract search effort
without changing the decoding objective. The BP marginal probabilities
guide the search through trust coefficients, and selected events of the
BP hard decision initialize a residual search. A signed-cost objective
allows the residual search to remove incorrect initialized events while
keeping the same channel-prior objective as Tesseract, and a
node-exploration limit with fallback bounds the additional search
effort when the initialization is poor.}

\textcolor{black}{Simulations on three BB codes under circuit-level noise show that
the proposed decoder reduces the number of expanded nodes and the
priority-queue size by up to more than two orders of magnitude at low
physical error rates, while keeping logical error rates close to those
of Tesseract. BP-based initialization with the signed-cost objective
reduces the search effort beyond BP guidance without a noticeable loss
of accuracy, and a global trust coefficient is sufficient, which
keeps the decoder simple to configure.}

Future work could explore extensions of \textcolor{black}{the proposed method to the
case of} correlated detector error models, particularly those coupling
$X$- and $Z$-sector events. Another direction is to combine the search
with alternative BP preprocessors, such as Relay BP~\cite{muller2025relaybp},
and study how their message-passing dynamics interact with Tesseract's
search. Reliability measures that identify incorrect BP decisions whose
\new{marginal probabilities} have stabilized would also benefit both
the \textcolor{black}{trust coefficients} and the selection of the initialization
set. \textcolor{black}{Finally, since BP is well suited to parallel FPGA
implementations~\cite{valls2021syndrome}, whereas the Tesseract search
relies on irregular priority-queue operations, a heterogeneous FPGA--CPU
implementation of the proposed decoder is a natural direction toward
real-time decoding.} 
\section*{Acknowledgment}
The authors thank Lamia Yous for helpful discussions and for sharing the code of~\cite{yous2026accelerating}.
\let\emph\textit
\bibliographystyle{IEEEtran}
\bibliography{reference}

\end{document}